\documentstyle[aps]{revtex}

\begin{document}
\title{Rate equations for cluster formation in supercooled liquids.}
\author{V. Halpern}
\address{Department of Physics, Bar-Ilan University, 52900 Ramat-Gan, Israel%
\\
E-mail: halpeh@mail.biu.ac.il}
\maketitle

\begin{abstract}
The formation of clusters in supercooled liquids close to the glass
transition temperature is described by rate equations in which the
coefficients are determined on physical grounds rather than in terms of
thermodynamic quantities such as free energies and surface tensions. In
particular, the density of free molecules in the liquid as a function of
temperature is determined self-consistently. Calculations for a very simple
model indicate that such rate equations are capable of producing  physically
reasonable results. Our results suggest that the difference between strong
and fragile liquids may be associated with the strength of the binding of a
surface molecule to a cluster, and they also provide indications about the
nature of the glass transition and the structure of the resulting glass.
\end{abstract}

\section{Introduction}

The slowing down of molecular rearrangement processes in supercooled liquids
as the temperature is reduced, until the system effectively freezes into a
glass (i.e. a non-crystalline solid) at the glass transition temperature $%
T_{g}$, is a phenomenon common to a very wide range of materials, from the
common silicate glasses used in windows to natural and artificial polymers.
The variety of unusual features found experimentally in these materials have
been the subject of several recent reviews \cite{Ediger}\cite{Angell-1}\cite%
{Angell-JAP}, as have the variety of theories proposed to account for them %
\cite{Silescu}. Most of these theories are of a general nature, because they
have to account for features observed in so many very different materials.
One of the most successful of them, which only assumes the existence of
non-linear interactions between density fluctuations, is mode coupling
theory (MCT) \cite{Gotze}. However, the original version of MCT ignores the
possibility of thermally excited processes, and predicts that the system
undergoes a phase transition into a frozen state at a temperature $T_{x}$
that is much higher than $T_{g}$. While MCT has been extended to include
thermally activated processes that enable this freezing to be avoided, and
so to describe the behavior of the system at temperatures between $T_{x}$
and $T_{g}$, these extensions involve specific assumptions about the types
of activated process and so are no longer so general. Numerous other types
of general theory have been advanced to describe the behavior of supercooled
liquids in this temperature range. Most of these involve the use of
thermodynamics, which is quite natural since this is a general framework,
and in particular because one of the main unusual features of the glass
transition is the behavior of the system's entropy as the liquid is cooled
towards $T_{g}$ \cite{Angell-1}. However, a major problem in many of these
theories is to relate the assumptions that they use to the molecular
structure of the different supercooled liquids and the microscopic
properties associated with them.

In this paper we describe an alternative approach to the problem of the
glass transition, in terms of the rate equations for the formation of
solid-like clusters of molecules in the supercooled liquid. There is
considerable experimental evidence for the existence of such clusters, which
cause the system to be inhomogeneous on certain length and time scales \cite%
{Richert}, and they are also postulated in many theoretical approaches \cite%
{Silescu}. However, not much attention is usually paid to the dynamics of
the formation and dissociation of such clusters, which in general can be
described .by rate equations. Rate equations are well known in the theory of
the nucleation of liquid-like droplets in a vapor and of crystallites in a
normal liquid,.where it is usually described by classical nucleation theory %
\cite{CNT}\cite{Olson}. It is also widely used in the theory of the
nucleation and growth of thin films on a substrate \cite{Family}.\ In both
these cases, the calculation of the rate constants for the growth and decay
of clusters involves well-defined thermodynamic concepts such an expression
for the difference in the free energy between the two phases, their chemical
potentials and the surface tension of the interfaces between them. A number
of authors, including Xia and Zinke-Allmang \cite{Xia-All} henceforth
referred to as XZ, have used a similar approach involving free energies and
surface tensions. However, in a supercooled liquid it may not be possible to
define such properties, or even if one does define them to relate them to
those of the crystalline solid and the normal liquid. One problem is that
the temperature dependence of the density of free molecules (which
presumably correspond to those of a normal liquid) is determined by the
self-consistent solution of the rate equations, which makes the definition
of the chemical potential of the liquid quite problematic. This point is
totally ignored in classical nucleation theory, and is not relevant for film
growth on a substrate where the free molecules (or atoms) are deposited at a
constant rate. Another possible problem with the use of these thermodynamic
quantities is the very irregular shapes of the clusters, as found in
molecular dynamics simulations \cite{Dzugutov}, which certainly do not
correspond to minima of the free energy in a classical picture. Accordingly,
we formulate here a very general approach to the rate equations for the
growth and dissociation of such clusters, and consider some of their
consequences. In principle, such a theory can describe not only the
structural properties of the system but also its dynamic properties, such as
the frequently observed non-exponential relaxation with time of correlation
functions, the dielectric response, and similar features. However, in the
initial analysis presented in this paper, we consider mainly the structural
properties.

The starting point of our approach is to postulate growth and decay rates
for clusters of molecules, without relating them in advance to the
thermodynamic potentials, and examine how their behavior influences the
properties of the system. In this paper, for the sake of simplicity we
restrict our analysis to the growth and decay of clusters one molecule at a
time. The rate of attachment of molecules to clusters is assumed to be
proportional to the density of single molecules, which it is convenient to
refer to as monomers. This is essentially a mean field approximation, and
ignores the possible existence of depletion zones around the clusters \cite%
{Family} and the effects of density fluctuations. The process of detachment
of monomers from a cluster is thermally activated, involving the breaking of
the bonds between a molecule on the surface of a cluster and the other
molecules of the cluster. A key novel feature of our approach, as noted
above, is that the monomer density (which is determined by or determines the
chemical potential of the liquid), is calculated self-consistently rather
than postulated. In section 2, we present the rate equations for the system
for discrete cluster sizes, and their solution for the steady state of an
extremely simple model. Such a steady state involves the dependence of the
monomer concentration and that of clusters of different sizes on the
attachment and detachment parameters, and so the temperature dependence of
these provides valuable information about how the steady state properties of
the system change with temperature. In the process of our analysis, a
fundamental difficulty is discovered in the continuum approximation used by
XZ \cite{Xia-All}\ to treat large clusters, which is discussed in the
Appendix. The results of our calculations are presented in section 3, and in
section 4 we discuss their significance and implications. In particular, our
analysis naturally leads to the model of glass structure recently proposed
by Stachurski \cite{Stach} of a ``maximum random jamming'' state, with small
number of ``rattler'' particles between them, rather than random close
packing to describe structures of ideal amorphous solids. A summary of our
results and conclusions is presented in section 5.

\section{The rate equations and their steady state solution.}

In a supercooled liquid, the total concentration of molecules is constant,
in contrast to the situation for thin film growth. Hence the most useful
form of the rate equations seems to be that of XZ \cite{Xia-All}, and we use
their formulation, but with a somewhat different notation. Let $n_{0}$ be
the total density of molecules, $n_{j}$ the density of clusters of $j$
molecules, and let $A_{j}n_{1}$ be the rate of single molecule attachment to
and $R_{j}$ the rate of detachment or release of single molecules from a
cluster of  $j$ molecules. If such clusters can have a variety of shapes,
the coefficients $A_{j}$ and $R_{j}$ are suitably weighted averages. Then
the basic rate equations are 
\begin{equation}
dn_{1}/dt=-2A_{1}n_{1}^{2}+R_{2}n_{2}+%
\sum_{j=2}^{n_{0}}(R_{j}-A_{j}n_{1})n_{j},
\end{equation}%
where the term $R_{2}n_{2}$ arises from fact that dissociation of a dimer
produces 2 monomers, while the dissociation of a larger cluster produces
only one monomer, and%
\begin{equation}
dn_{j}/dt=n_{1}(A_{j-1}n_{j-1}-A_{j}n_{j})+(R_{j+1}n_{j+1}-R_{j}n_{j})
\end{equation}%
In addition, in our mean field approximation the total density of molecules
in the system is fixed, so that%
\begin{equation}
\sum_{j=1}^{n_{0}}jn_{j}=n_{0}
\end{equation}%
In view of this equation, equation (1) must follow from the sum for $j>1$ of 
$j(dn_{j}/dt)$\ as given by equation (2), a point that is readily checked
and makes it unnecessary to use equation (1).

While the continuum approximation discussed in the Appendix may be needed
for analyzing the growth and decay of clusters, for the steady state
solution it is simpler (and also more accurate) to use the exact equations
(2)-(3). In order to obtain qualitative ideas of how the solution behaves,
we consider the case where $A_{j}$ and $R_{j}$ both depend on $j$ only
through the surface area $g_{j}$ of a cluster of $j$ molecules (which could
be a reasonable approximation for large clusters) and ignore the fact that
clusters containing the same number of molecules may have different shapes
and so different surface areas and binding energies of the surface
molecules. Thus, we write%
\begin{equation}
A_{j}=A_{0}g_{j},\quad R_{j}=R_{0}g_{j},\quad j\geq j_{0}.
\end{equation}%
It is convenient to write%
\begin{equation}
g_{j}n_{j}=f_{j},
\end{equation}%
so that equation (2) for the steady state can be written in the form%
\begin{equation}
n_{1}A_{0}(f_{j-1}-f_{j})+R_{0}(f_{j+1}-f_{j})=0,\quad j\geq j_{0}+1
\end{equation}%
Equation (6) is a simple second order linear difference equation for $f_{j}$%
, the general solution of which is $f_{j}=c_{1}z_{1}^{j}+c_{2}z_{2}^{j}$,
where $z_{1}$ and $z_{2}$are the roots of 
\begin{equation}
R_{0}z^{2}-(R_{0}+n_{1}A_{0})z+n_{1}A_{0}=(z-1)(R_{0}z-n_{1}A_{0})=0
\end{equation}%
The root $z=1$ leads to $n_{j}=c/g_{j}$, which leads to the divergence of
the sum in equation (3) unless $\lim_{j\rightarrow \infty }(g_{j}/j^{2})>0$,
and this is obviously impossible. Hence the only solution of interest is 
\begin{equation}
f_{j}=cz_{0}^{j},\quad z_{0}=(A_{0}/R_{0})n_{1}\equiv bn_{1}
\end{equation}%
For convenience, we assume that 
\begin{equation}
g_{j}=B_{0}j^{\alpha },
\end{equation}
and extend the sum in equation (3) to $\infty $, which is justified since
this sum then converges for $0<z_{0}<1$ according to the ratio test. In view
of the definition of $g_{j}$ in equation (4), we can choose $B_{0}=1$.

In the simplest (but totally unrealistic) case that equation (4) holds for
all $j>1$, it follows from equation (3) that $z_{0}=bn_{1}$ is the root of
the equation 
\begin{equation}
c\sum_{j=1}^{\infty }j^{1-\alpha }z_{0}^{j}=n_{0},
\end{equation}%
subject to the condition that $z_{0}<1$, i.e. $bn_{1}<1$ Moreover, since by
definition the number of isolated molecules is also the number of clusters
of single molecules, $n_{1}=f_{1}=cz_{0}=c(bn_{1})$, and so $c=1/b$. Hence,
finally, in this case $n_{1}$ is the root of the transcendental equation%
\begin{equation}
\sum_{j=1}^{\infty }j^{1-\alpha }(bn_{1})^{j}=bn_{0}
\end{equation}

Intuitively, one expects that the exact values of $A_{j}$ and $R_{j}$ for
small values of $j$ will not affect the qualitative behavior of the results.
In order to test this hypothesis, we performed calculations for the above
system, which we call system 1, and for system 2 in which equation (4) is
not assumed to be valid for dimers, so that $j_{0}=3$. Since the thermally
activated dissociation of a dimer only involves the breaking of one bond,
while the release of a molecule from a trimer usually involves the breaking
of two bonds, we chose the activation energy of $R_{2}$ to be half that of $%
R_{0},$and wrote $A_{2}/R_{2}=\surd (A_{0}/R_{0})$, while since a monomer
has only one site for attachment and a dimer has two sites we chose $%
A_{1}=A_{0}$ and $A_{2}=2A_{0}$. After simple calculations, we then find
that $n_{j}=c(bn_{1})^{j}$ for $j\geq 3$, $n_{2}=(%
{\frac12}%
\surd b)n_{1}^{2}$, and $c=b^{-1.5}$, and also adjust accordingly the first
two terms in the sum in equation (11).

\section{Results of the calculations}

In order to understand the physical significance of the results of our
calculations, before presenting them we consider the temperature dependence
of the parameters $A_{0}$ and $R_{0}$, and hence of the parameter $b$, which
determines the value of $n_{j}/n_{0}$. The release of a molecule from a
cluster requires the breaking of bonds between it and the remaining
molecules in the cluster, and so is expected to be a thermally activated
process. Hence $R_{0}$ should depend exponentially on the temperature $T$,
and we write $R_{0}=R_{00}\exp [-E_{a}/(kT)]$. The addition of a molecule to
a cluster, on the other hand, will often be controlled by diffusion \cite%
{Olson}, with a very weak intrinsic temperature dependence (although it may
well be affected by the temperature-dependent cluster density), and even if
it does require some thermal activation the energy required is far less than
that required to detach a molecule from the cluster. Thus, to the order of
approximation inherent in our very simple model systems, we can assume that 
\begin{equation}
b=b_{0}\exp [E_{0}/(kT)],
\end{equation}%
with a positive activation energy $E_{0}$, so that an increase in the
parameter $b$ corresponds to a decrease in the temperature $T$.

The results that we report are all for $n_{0}=1,$and for $\alpha =2/3$,
which is the appropriate value for spheres in three dimensions; the exact
value of $\alpha $ does not affect the qualitative behavior of the results.
In figure 1, we show the value of $n_{1}$ as a function of $\log (b)$ for
the two systems.. The main point to notice is that for both systems $n_{1}$
decreases as $b$ increases, which corresponds to a decrease in the
concentration of monomers as the temperature is lowered. For system 2, the
value of $n_{1}$ is lower than in system 1 for $b<1$ and higher than in
system 1 for $b>1$,because here $A_{2}/R_{2}=\surd (A_{0}/R_{0})=\surd b$,
which is larger than $b$ for $b<1$ and less than $b$ for $b>1$, while as
noted above an increase in $b$ leads to a decrease in $n_{1}$. The
difference is greatest for the low values of $b$ because in this region
dimers are of greater importance. As can be seen clearly in figure 2, in
which the results of figure 1 are replotted on a double logarithmic scale,
the decrease in $\log (n_{1})$ with increasing $b$ for $b>5$ is linear in $%
\log (b)$, so that in view of equation (12) in this region $n_{1}=n_{10}\exp
[-E_{1}/(kT)]$, and a least squares fit shows that $E_{1}$ is virtually the
same as $E_{0}$. For both systems we find that this exact exponential
decrease occurs when $n_{1}<0.2$, while the decrease is nearly exponential
from $b=1$, where $n_{1}=0.46$ and $0.47$\ for systems 1 and 2 respectively.
For lower values of $b$, $n_{1}$ decreases much more slowly as $1/T$
increases.

A much more interesting difference between systems 1 and 2 is with regard to
the cluster size $j_{\max }$\ containing the largest number of molecules..
The value of $j_{\max }$ is given by the maximum of $jn_{j}$, i.e. of $%
j^{1-\alpha }z_{0}^{j}$, which occurs at $j=(1-\alpha )/\ln (1/z_{0})$,
where $z_{0}=bn_{1}$.\ We find that for system 1 $j_{\max }=3$ when $b=14$
and $n_{1}=0.065$, and $j_{\max }=10$ when $b=80$ and $n_{1}=0.012$, while
for system 2 the corresponding values are $b=6.3,$ $n_{1}=0.14$ and $b=20,$ $%
n_{1}=0.05$ respectively. While in both cases the values of $n_{1}$ are much
too small to be realistic, the appearance of clusters of a given size at
appreciably larger values of $n_{1}$ for system 2 than for system 1
indicates that our general physical picture is reasonable, so that more
realistic descriptions of the size dependence of the accretion and removal
rates of particles should lead to physically meaningful results.

Finally, we consider the temperature dependence of the system's
configurational entropy.. For monomers, this should correspond to that $S_{l}
$ of molecules in the liquid, while for molecules within clusters it should
as a first approximation be zero, as for molecules in the solid, but for
molecules on the surface of a cluster it should have an intermediate value $%
S_{surf}$. Accordingly, we write for model 1 
\begin{equation}
S=n_{1}S_{l}+a(\sum_{j=2}^{\infty }j^{\alpha }n_{j})S_{surf}
\end{equation}%
where $a$ is a geometrical factor. Since $j^{\alpha }n_{j}=f_{j}=z_{0}^{j}/b,
$it follows that 
\begin{equation}
S=n_{1}S_{l}+aS_{surf}z_{0}^{2}/(b-bz_{0})
\end{equation}%
In figure 3 we show the contribution to the configurational entropy from the
molecules on the surfaces of the clusters, and the total entropy for the
arbitrary choice of $aS_{surf}=S_{l}/3$, as a function of $\log (b)=\log
(b_{0})+E_{0}\log (e)/(kT)$. As can be seen, the decrease of the
configurational entropy with decreasing temperature (increasing $b$) is
qualitatively similar to that which is observed experimentally. Also, at low
temperatures (large $b$) the main contribution to this entropy comes not
from the small number of monomers but rather from the molecules on the
surfaces of the clusters.

\section{Discussion}

While the model presented above is obviously far too simple to represent any
real system, it does indicate various trends that are physically plausible
and worthy of detailed investigation using suitable extensions of our
models. The first of these regards the approach to the glass transition in
supercooled liquids. Here, our results show the crucial role played by the
parameter $b$, which is the ratio of the attachment rate to a cluster per
monomer around it to the rate of detachment of the molecule from the
cluster. We found that for small values of $b$ the monomer concentration is
not very sensitive to its value, but for large values of $b$ it decreases
rapidly as $b$ increases. Physically, the reason for this is that large
values of $b$ correspond to the molecules on the surface of a cluster being
strongly bound to it, so that the formation of clusters permanently depletes
the monomer population, and the temperature dependence of the detachment
rate dominates the value of $b$. For weakly attached surface molecules, on
the other hand, the reduction of the diffusion rate of monomers as the
temperature is lowered and the cluster density increases will\ have a much
larger effect on the temperature dependence of the parameter $b$, which will
become non-exponential as a result. Since a non-Arrhenius temperature
dependence is typical of \ fragile glasses, this suggests that the
distinction between strong and fragile glasses may be related to whether the
molecules on the surface of a cluster are strongly or weakly bound to the
cluster.

While our model does not directly consider the viscosity of the system,
which is a dynamic rather than a structural property, since it does
determine the cluster density it could be used to determine the viscosity 
in conjunction with cluster models for this, such as that of Fan and Fecht %
\cite{Fan}, without making their assumptions about the free energies of the
clusters as they do. Similarly, our model can account for the observed rapid
transition of particles from fast states, corresponding to free molecules,
to slow states corresponding to bound ones that is observed experimentally %
\cite{Richert} \ This transition also has a strong effect on the dielectric
response of the system, since in a correct treatment of the dielectric
response and relaxation functions \cite{Halpern} the greater response of
free molecules to an applied field plays an important role.

Finally, our analysis strongly supports the models of Bakai \cite{Bakai}\
and Stachurski \cite{Stach}, without their assumptions about the
thermodynamic potentials, that the transition from a supercooled liquid to a
glass occurs when the solid clusters coalesce or combine with each other and
jam, forming a system in which free molecules can no longer percolate. Such
coalescence can also explain the strange shapes of clusters found in
molecular dynamics simulations \cite{Dzugutov}. While the coalescence of
clusters is of vital importance for properties of the system such as
viscosity, correlation lengths and slow modes, such a coalescence as a
result of the proximity of a pair clusters or the growth or a bridge between
them will often only lead to only a small change in the attachment and
detachment rates of monomers, so that it should be sufficient to use the
rate equations for small clusters

\section{Conclusions}

In this paper, we proposed that the densities of clusters of different sizes
in a supercooled liquid be calculated from the steady state solution of rate
equations rather than from postulated values of the thermodynamic potentials
or free energies and the surface tensions. The advantage of such an approach
is that the values of the attachment and detachment coefficients appearing
in the rate equations can be given a simple physical significance, and it
should be possible to relate them to the elementary properties of the
molecules in the system. Our calculations for a very simple model shows that
the rate equations are capable of giving qualitatively reasonable results,
so that it is worthwhile to use them to study more sophisticated models and
time-dependent properties. One consequence of our approach is that it is not
necessary to assume that large cooperatively rearranging regions appear as
an entirely new phenomenon near the glass transition temperature, since the
temporary coalescence of small clusters can give rise to such domains.
Moreover, according to this approach the glass transition occurs when the
density of clusters is so large that they can no longer move independently,
i.e. when their motion is jammed.

\section{Appendix - The continuum approximation}

For treating clusters containing a large number of molecules in supercooled
liquids, XZ \cite{Xia-All}\ proposed replacing the discrete index $j$ for $%
j>2$ by the continuous variable $x$, and replacing the second order
difference equation (3) in $j$ by a first order differential equation in $x$%
, 
\begin{equation}
\partial n(x,t)/\partial t=-n_{1}(t)[\partial /\partial
x][A(x)n(x,t)]+[[\partial /\partial x][R(x)n(x,t)]  \eqnum{A1}
\end{equation}%
A basic problem with the above continuum approximation is that in the steady
state equation (A1) becomes 
\begin{equation}
n_{1}(d/dx)[A(x)n(x)]=(d/dx)[D(x)n(x)],\quad x\geq 3  \eqnum{A2}
\end{equation}%
The solution of this equation is $n_{1}A(x)n(x)=D(x)n(x)+c$, and since $%
n(x)\rightarrow 0$ as $x\rightarrow \infty $ one expects that $c=0$, in
which case $n_{1}A(x)=D(x)$\ This solution corresponds to the root $z=1$ in
our exact analysis. which as we saw is physically untenable. Hence, a better
continuum approximation is required, and this can be obtained by writing
down the difference between equation (2) for $j$ and $j+1$%
\begin{equation}
(d/dt)(n_{j+1}-n_{j})=-n_{1}(A_{j=1}n_{j+1}+A_{j-1}n_{j-1}-2A_{j}n_{j})+(D_{j+2}n_{j+2}-D_{j}n_{j}-2D_{j+1}n_{j+1})
\eqnum{A6}
\end{equation}%
and then applying the continuum approximation \ For the time-dependent
equation it gives a non-linear partial differential equation of second order
in $x$ and first order in $t$, which has to be solved together with an
equation for the conservation of particles which involves integrals of the
solution. For the steady state, the continuum equation corresponding to this
is%
\begin{equation}
n_{1}(d^{2}/dx^{2})[A(x)n(x)]=(d^{2}/dx^{2})[D(x)n(x)],\quad x\geq 3 
\eqnum{A7}
\end{equation}%
with a solution $n_{1}(d/dx)[A(x)n(x)]=(d/dx)[D(x)n(x)]+c$, which is more
promising, even if $c=0$ for the same reasons as previously, since its
solution need not contradict the equation for particle conservation. In fact
this solution corresponds to the root $z_{0}=(A_{0}/R_{0})n_{1}\equiv bn_{1}$%
.in our exact analysis, which as we saw was the physically correct one.
Hence the approximation of XZ of replacing the exact rate equations by
continuum equations involving only first derivatives in the cluster size
does not seem to be justified.

\bigskip

Captions for figures

Figure 1: \ The density $n_{1}$ of free molecules (monomers) as a function
of $\log (b)$. The full curve is for system 1, and the broken curve for
system 2

Figure 2. The logarithm of the density of free molecules (monomers) $\log
(n_{1}$ ) as a function of $\log (b)$. The full curve is for system 1, and
the broken curve for system 2

Figure 3. The entropy $S$ of the system, for $S_{surf}=S_{1}/3$ (full curve)
and $S_{surf}$ (broken curve) as functions of $\log (b)$.

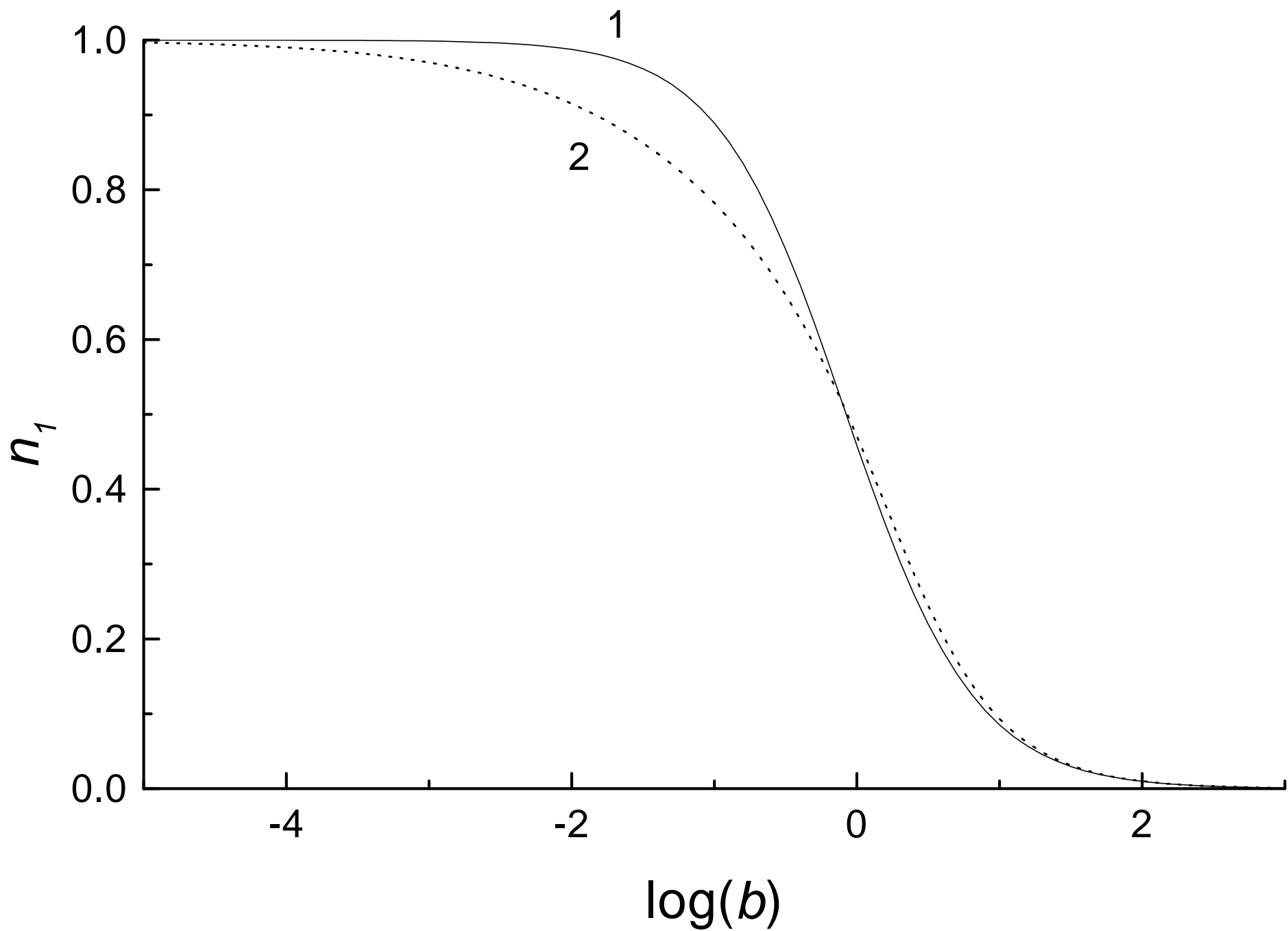
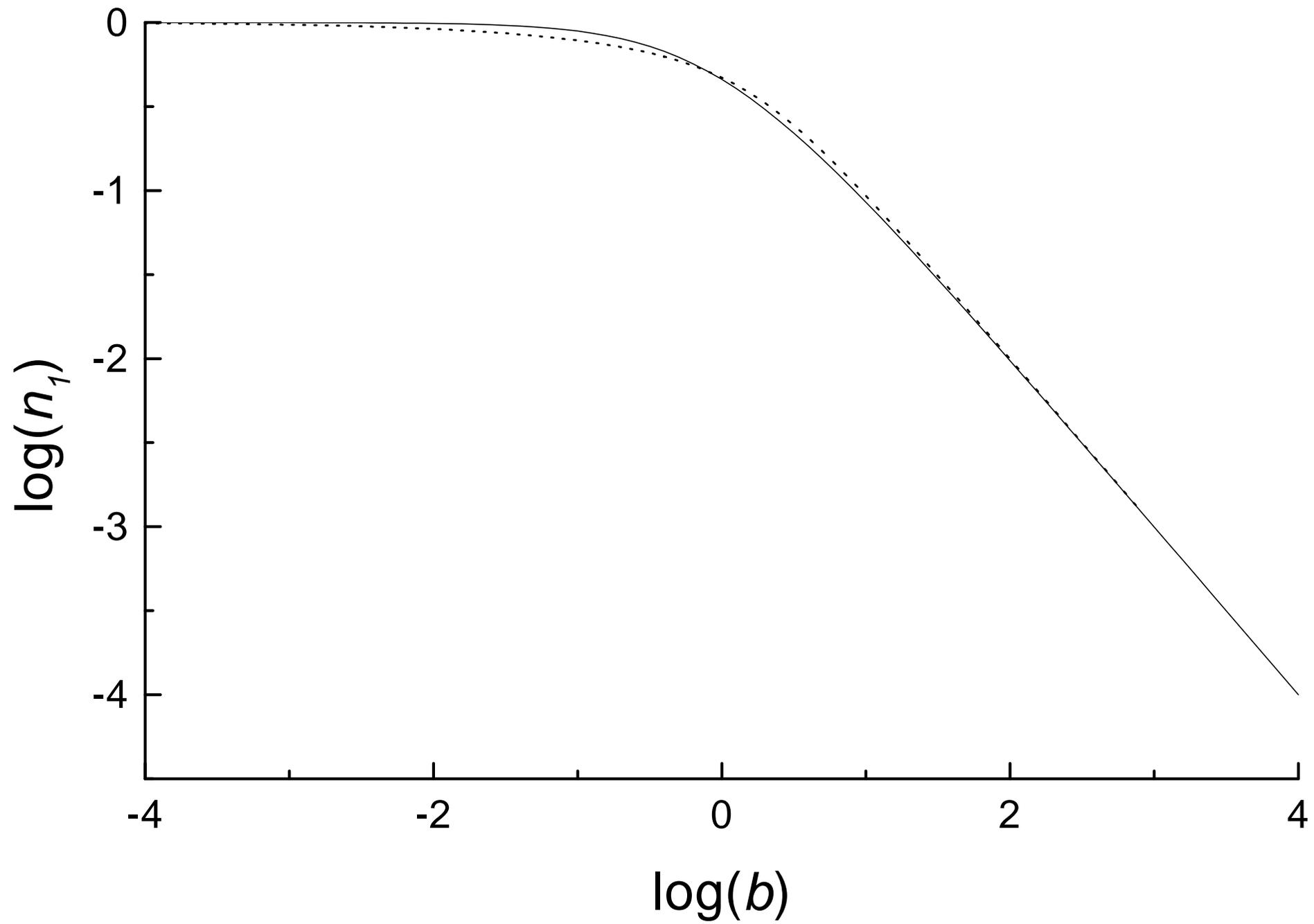
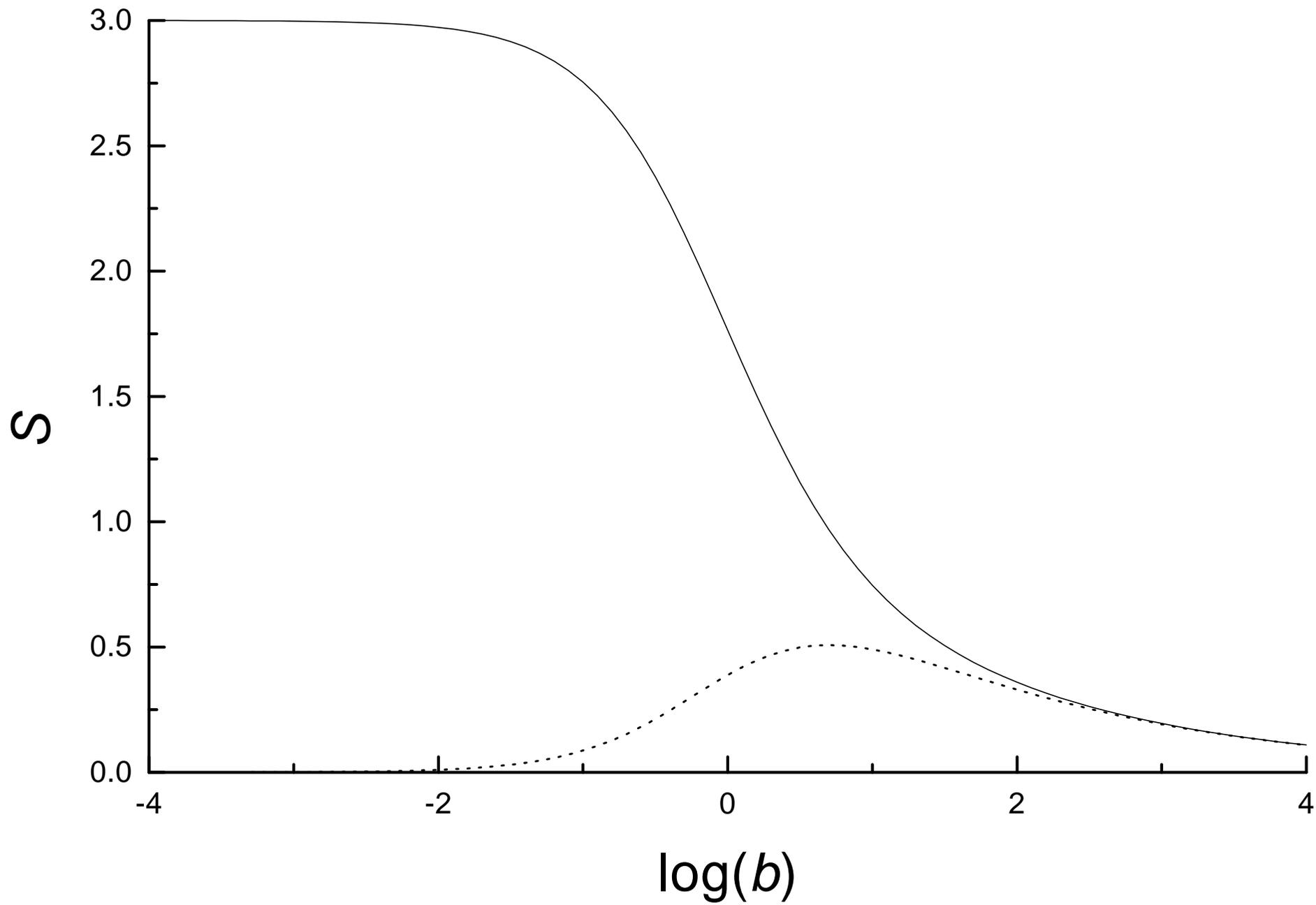
\end{document}